\documentclass{acm_proc_article-sp}
\usepackage{subcaption}
\usepackage[numbers]{natbib}
\usepackage[ruled]{algorithm2e}
\usepackage{epstopdf} 
\DeclareTextFontCommand{\emph}{\bfseries \itshape}
\usepackage{epsfig} 
\makeatletter
\def\@copyrightspace{\relax}
\makeatother

\author{\vspace{-12pt}
}

\title{ Extract Secrets from Wireless Channel: A New Shape-based Approach} 

\begin{document}  
         
  \maketitle
  \subsection*{ABSTRACT}
  \vspace{5pt}

Existing secret key extraction techniques use quantization to map wireless
channel amplitudes to secret bits. This paper shows that such techniques are
highly prone to environment and local noise effects: They have very
high mismatch rates between the two nodes that measure the channel
between them. This paper advocates using the shape of the channel instead of
the size (or amplitude) of the channel. It shows that this new paradigm shift is
significantly robust against environmental and local noises. We refer to this
shape-based technique as {\bf Puzzle}. Implementation in a software-defined
radio (SDR) platform demonstrates that Puzzle has a 63\% reduction in bit
mismatch rate than the state-of-art frequency domain approach (CSI-2bit).
Experiments also show that unlike the state-of-the-art received signal strength
(RSS)-based methods like ASBG, Puzzle is robust against an attack in which an
eavesdropper can predict the secret bits using planned movements.


  \section{Introduction}\label{introduction}
\vspace{5pt}

For wireless communications, there has been a great interest in generating shared secrets from the physical layer as a complementary approach to the traditional methods of cryptography. 
The interest stems from the open nature of the wireless medium and the infrastructure constraints associated with key management in mobile scenarios. There are two main approaches for 
secret-sharing in wireless. One is based on information-theoretic principles of exploiting the secrecy capacity between Alice and Bob compared to Alice and Eve \cite{erasureChannel}. The 
main drawback of this approach is that secrecy is dependent on rather strong assumptions about eavesdropper capability.  
Equally importantly, even a modest increase in the spatial density of eavesdroppers harms the secrecy rate of the approach dramatically \cite{MultiEves}.

The other approach is based on channel reciprocity. Channel reciprocity refers to the physical principle whereby near-simultaneous observations of the channel by two communicating parties 
are identical due to the channel paths between them being symmetrical. Figure~\ref{fig:smoothed} in Section~\ref{generation} shows this reciprocity in our testbed. The time for which the wireless channel remains correlated is called the \emph{coherence time}. By extracting channel 
state information from the observed signals, Alice and Bob can share bits by transmitting signals to each other within the coherence time. Furthermore, extensive theoretical analysis and experimentation have 
shown that observations of the wireless channel over distances larger than half-the-wavelength of the carrier frequency are uncorrelated \cite{Rappaport:2001:WCP:559977}. In a 2.4GHz ISM 
band, for instance, at any location farther than 6cm away from Bob, Eve will observe Alice's signal through an uncorrelated channel. Channel reciprocity and spatial decorrelation together make 
the wireless channel an excellent random source for generating shared secret keys.

There is significant prior work that exploits channel reciprocity for secret extraction. One set of techniques use the received signal strength (RSS) as the secret source 
\cite{Azimi-Sadjadi:2007:RKG:1315245.1315295, ASBG,Mathur:2008:RES:1409944.1409960, Patwari:2010:HUB:1687041.1687073, multi-antenna}. These techniques measure the received signal 
strength over different coherent times to generate a sequence of received signal strengths. They choose a threshold and transform the signal strength sequence into 1s (if above that threshold) and 
0s (if below the threshold). The largest drawback with RSS-based techniques is that large variations can be easily introduced by an 
attacker by blocking transmission every now and then. These make the secret predictable since the attacker knows the exact moments at which the signal-to-noise ratio (SNR) will drop or increase. 
Section~\ref{validation} presents this attack and shows this vulnerability. Even if there are no malicious attackers, some unintentional regular activities would also make the variation public. For example, SNR 
in a corridor of a classroom building would be much lower after class than during class. 


  		
Another set of techniques use the fine-grain temporal
\cite{Wilson07channelidentification, MadisehMNCH08,
Madiseh:2009:VSK:1817271.1817382, Ye_onthe} or frequency \cite{freq} components
contained in received signals as the secret source.  The temporal techniques
use ultra-wideband transmissions ($\approx$ GHz bandwidth) to capture this
fine-grained temporal information. Therefore, these techniques are not
applicable for narrowband systes such as Wi-Fi (with only 20MHz bandwidth).
Furthermore, another challenge in temporal techniques is that temporal
information is sensitive to sampling offset which leads to a high rate of
secret disagreement.  In contrast, the frequency technique of Liu et al
\cite{freq} is applicable to narrowband systems and is not sensitive to 
sampling offset. The authors quantize the
frequency response in each subcarrier in OFDM and map them to 
secret bits. In Section~\ref{validation}, we dispute the authors' claim 
of high secrecy rate and show that the secrets generated from their method 
is very limited.

Overall, this paper takes the stand that the amplitude (the {\bf size}) of a 
signal --in time or frequency-- is prone to perturbations from the environment 
as well as hardware imperfections. This leads to quantization errors at nodes 
and high mismatch in secrecy bits generated by wireless nodes. Instead, this 
paper proposes to use the {\bf\emph{shape}} of a signal to deduce secrecy bits. 
Specifically, we make the following contributions.
\begin{itemize}
\item We propose and implement a shape-based secret extracting algorithm called Puzzle that we show to be robust to noise and device imperfections. In particular, no online or offline device calibration is required in using our algorithm.
\item We prove that the power spectrum density (PSD) of random data can be used to extract the channel state information. This implies that no modification is needed for the higher layers of the wireless communication, such as transmitting special training data. Two communicating parties can successfully extract secrets from the received packets as long as they exchange data packets within their coherence time.
\item Our experiments show that Puzzle produces a 5-bit secret per packet and has a 63\% improvement in bit mismatch rate than the frequency domain approach mentioned above. 
\end{itemize}

  \section{System Model}\label{model}
Consider two wireless nodes, Alice and Bob, that wish to create a shared secret $S$ within a coherence time, during which the channel is stable. 
An adversary, Eve eavesdrops the communication between Alice and Bob. Our goal is to develop a secret extraction algorithm that introduces as little communication and computation overhead as possible and ensures that Eve obtains little information about $S$. 
\subsection{Physical Layer Model}
	\subsubsection{Channel model}
	\setlength{\belowdisplayskip}{2pt} \setlength{\belowdisplayshortskip}{2pt}
	\setlength{\abovedisplayskip}{2pt} \setlength{\abovedisplayshortskip}{2pt}
Assume Alice and Bob operate in a Time-Division Duplexing (TDD) system. If they talk to each other in coherence time, the observed signals of Alice and Bob are represented by
	\begin{equation} \label{eq:ob_A}
	y_{A}(t)\ =\ (h*x_{A})(t)+n_{A}(t)
	\end{equation}
	\begin{equation} \label{eq:ob_B}
	y_{B}(t)\ =\ (h*x_{B})(t)+n_{B}(t)
	\end{equation}
where $h(t)$ is the channel impulse response, which is identical in both directions by virtue of channel reciprocity, $x_{A}$ and $x_{B}$ are the signals transmitted by Alice and Bob respectively, $n_{A}(t)$ and $n_{B}$ are additive white Gaussian noise with the same variance $N$, and ``$*$" indicates convolution. In the frequency domain, the equations above are rewritten as 
\begin{equation}\label{eq:Y_A'}
	Y_{A}(f)\ =\ H(f)\cdot X_{A}(f)+N_{A}(f),\ \frac{-W}{2}+f_{c}<f<\frac{W}{2}+f_{c}
	\end{equation}
	\begin{equation}\label{eq:Y_B'}
	Y_{B}(f)\ =\ H(f)\cdot X_{B}(f)+N_{B}(f),\ \frac{-W}{2}+f_{c}<f<\frac{W}{2}+f_{c}
	\end{equation}
where $W$ is the transmission bandwidth, $f_{c}$ is the center frequency, and $H(f)$ is the channel frequency response.	
\subsubsection{Channel Frequency Response}\label{sec:reciprocity}
In this section, we propose two ways to extract the channel frequency response $H(f)$.
\begin{itemize}
\item \textbf{Direct calculation}:
By using pre-defined training signals or decoding the received signals, Alice and Bob know the frequency components $X_{A}(f)$ and $X_{B}(f)$ of the transmitted signals. Therefore, they can calculate $H(f)$ easily, assuming that noise can be ignored. 
\item \textbf{PSD based method}:
Let $\{\ x_{0},x_{1},...,x_{N-1}\ \}$ be a complex sample sequence. Since the sequence is stationary and random, the auto-correlation of the sequence is
	\begin{equation}\label{eq:covariance}
	R(t_{1},t_{2})\ =\ \frac{P}{N}\times\delta(t_{2}-t_{1}) 
	\end{equation}
where $P$ is the power contained by the signal sequence.
Then, the PSD of the sequence is
	\begin{equation}\label{eq:fft}
	F[R(\tau)]\ =\ \int_{-\infty}^{+\infty}\frac{P}{N}\times\delta(\tau)e^{-j\omega\tau}d\tau = \frac{P}{N}
	\end{equation}
From Equation~\ref{eq:fft}, we know that 
	\begin{equation}\label{eq:X_A}
	X_{A}(f)\ =\ \frac{P_{A}}{W},\quad
	X_{B}(f)\ =\ \frac{P_{B}}{W}
	\end{equation}
Combining Equations \ref{eq:Y_A'} through \ref{eq:X_A} we get 
	\begin{equation}\label{eq:Y_A}
	Y_{A}(f)\ \approx\ \frac{H(f)\cdot P_{A}}{W}+N,\ 
	Y_{B}(f)\ \approx\ \frac{H(f)\cdot P_{B}}{W}+N
	\end{equation}	 

According to the above equations, we conclude that the PSD of $y_{A}(t)$ is the same as that of $y_{B}(t)$ as long as $P_{A}\ =\ P_{B}$. 
It is worth noting that even if $P_{A} \ \neq \ P_{B}$, the shape of Alice's and of Bob's PSD are still similar. This property is remarkable because it can be extended to the case in which Alice and Bob experience different levels of transmission power, noise or cross-band interference. Even in such cases, the shapes still don't change significantly. 
\end{itemize}
\subsection{Threat Model}
Eve is motivated to derive the shared secret generated by Alice and Bob. There are two main ways of achieving this.
\subsubsection{Eavesdropping}
Eve can attempt to derive $Ch_{AB}$ from $Ch_{AE}$ or $Ch_{BE}$, where $Ch_{AB}$, $Ch_{AE}$, and $Ch_{BE}$ denote the channel from Alice to Bob, Alice to Eve, and Bob to Eve, respectively. This may be possible if Eve has full knowledge of the environment. In general, however, full knowledge of the environment is a rather unrealistic assumption, so we do not regard it as the main threat to our system. Instead, we focus on the threat of spatial correlation of the secrets produced by our algorithm. We assume that Eve cannot stalk Alice or Bob to being within half of a wave length of either of them. This assumption is reasonable since close eavesdroppers suffer from a high exposure risk. Recall that theory \cite{Rappaport:2001:WCP:559977} supports that channels decorrelate beyond half a wavelength. 
\subsubsection{Planned movement}
Eve can move in between Alice and Bob to block and unblock their transmissions. Planned movements can thus introduce predictable increase or decrease of RSS at Alice and Bob. Note that while this attack is harmful to RSS-based methods, without the full knowledge about the environment, Eve cannot, however, predict the impact of the planned move on the frequency response of the channel.

 \section{Secret Generation}\label{generation}
After getting the frequency response curve of the received samples, we smoothen the curve, encode the smoothed curve by segmenting it into several pieces, and then map each of the pieces into one of the patterns from a predetermined set. 
\begin{figure}
\centering
	\begin{minipage}[t]{\columnwidth}
		\begin{subfigure}[t]{.49\columnwidth}
		\includegraphics[width = \columnwidth]{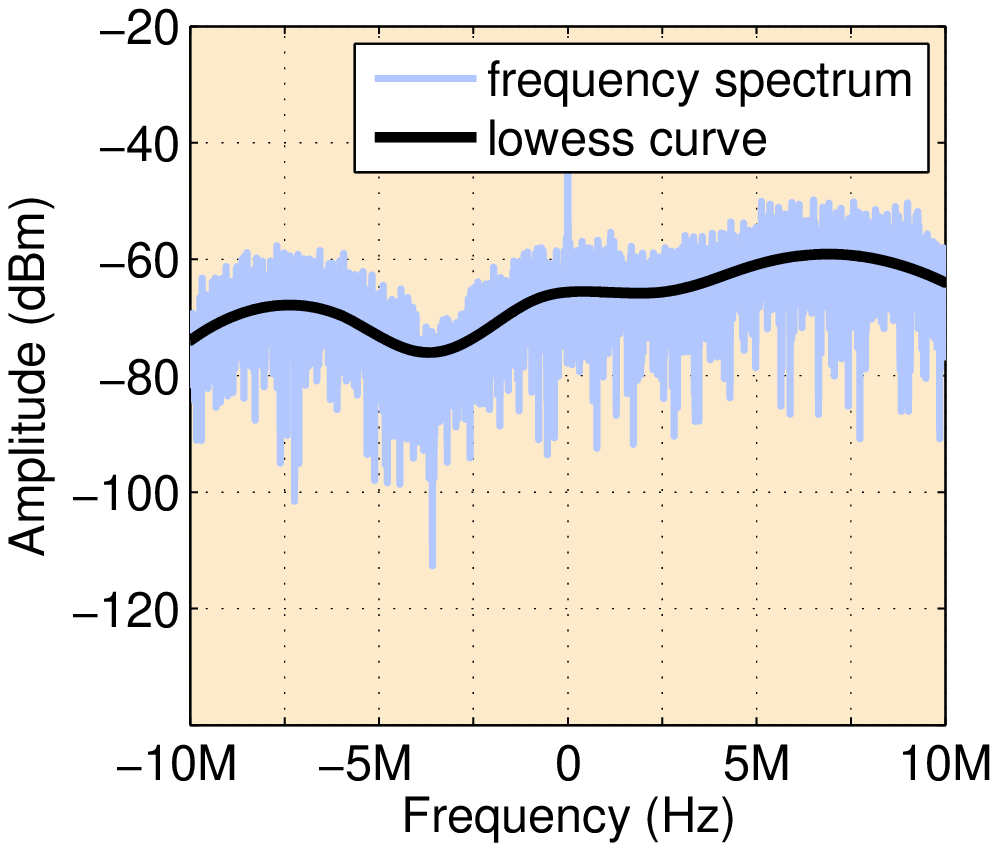}
		\caption{Lowess curve derived by Alice}
		\label{fig:smoothA}
	\end{subfigure}
	\begin{subfigure}[t]{0.49\columnwidth}
		\includegraphics[width=\columnwidth]{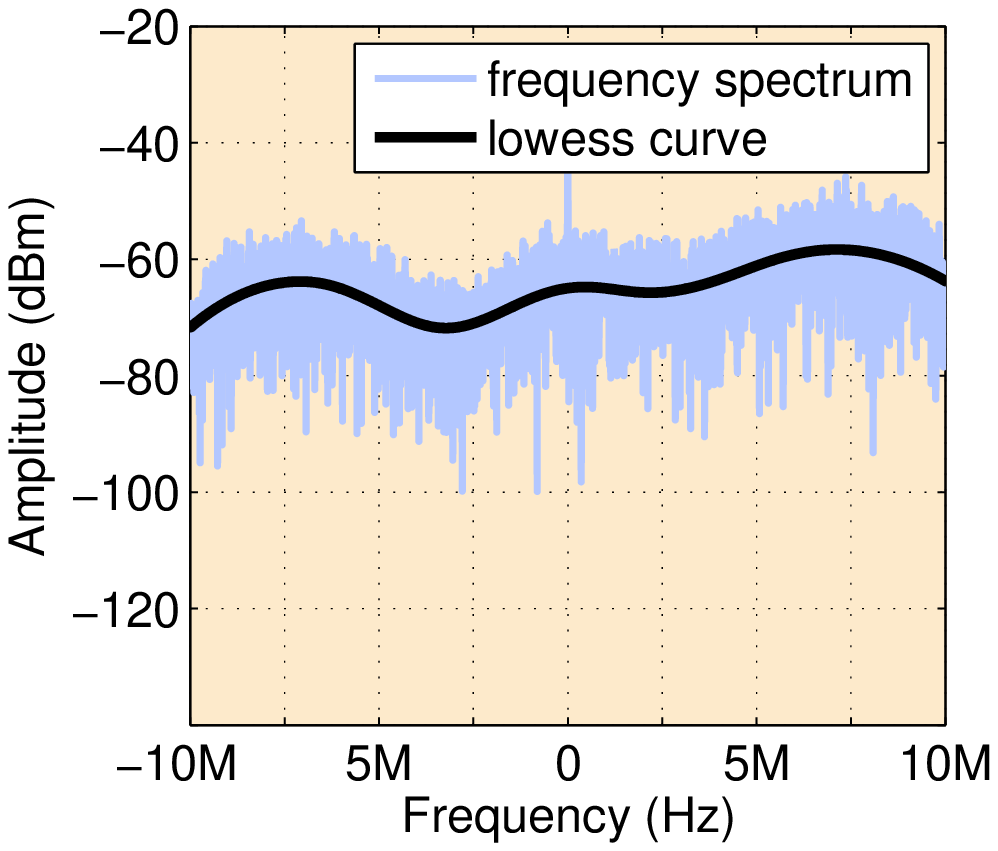} 
		\caption{Lowess curve derived by Bob}
		\label{fig:smoothingB}
	\end{subfigure}
	\caption{Lowess curves derived by Alice and Bob. Lowess curves are much more similar to each other than the original PSD curves as local variations are removed.}
	\label{fig:smoothed}
	\end{minipage}
\end{figure}
Note from Fig.~\ref{fig:smoothed} that although channel reciprocity is clearly apparent for the naked eye, the frequency response curves are more or less shifted or zoomed versions at corresponding frequencies. Moreover, distinct local fluctuations exist. These discrepancies are unavoidable because they spontaneously result from the hardware imperfections and environment interferences. This shows that direct quantization and mapping of the frequency response can lead to high mismatch rates. We, therefore, develop a shape-based approach to solve the encoding problem. 
\begin{algorithm}[h]
		\caption{CurveCoding}
		\KwIn{\\complex samples $a[0,\cdots,n]$\;number of segments $m$;}
		\KwOut{\\code $[C_{1},C_{2},\cdots,C_{m}]$}
		\SetKwFunction{Max}{Max1}
		\SetKwFunction{Min}{Min1}
		\SetKwFunction{PatternGeneration}{PatternGeneration}
		\SetKw{KwInit}{Initialization}
		\KwInit{
		\\divide $a[0,\cdots,n]$ into $m$ segments $b_{1}$,$b_{2}$,$\cdots$,$b_{m}$\;				 			$peak =\ \{\ $\Max{$a[0, \cdots, n]$} - \Min{$a[0, \cdots, n]$}$\}$		
		\PatternGeneration{$\lfloor n/m\rfloor$,$m$, $peak$}:\\
				\quad	generate 3 patterns of size $\lfloor n/m\rfloor$: $p_{1}$,$p_{2}$,$p_{3}$\;
					}					
		\For{$i\leftarrow 1$ \KwTo $m$}
				{$temp$ = $\infty$\;
				\For{$j = 1 \to 3$}
					{$dis$ = Fr\'echet($b_{i}$,$p_{j}$)\;
					\If{$temp$ $>$ $dis$}
						{$temp$ =  $dis$\;
						$C_{i}$ = $j$\;}
					}
				}
	\label{alg:coding}	
	\end{algorithm}
\begin{algorithm}[h]
			\caption{PatternGeneration}
			\KwIn{\\$k$, $m$, $peak$;}
			\KwOut{\\ 3 patterns $p_{1}[1,\cdots,k]$, $p_{2}[1,\cdots,k]$, $p_{3}[1,\cdots,k]$}
			\For{$i\leftarrow 1$ \KwTo $k$}
				{$p_{1}[i]\ =\ \frac{peak\ \times\ i}{k}$\;
				 $p_{2}[i]\ =\ -\frac{peak\ \times\ i}{k}$\;
				 $p_{3}[i]\ =\ \frac{peak}{m/2}$\;}
	\label{alg:pattern}
\end{algorithm}
\subsection{Curve Smoothing}\label{sec:curve_smoothing} 
As mentioned above, even though local details of a power spectral density pair are significantly different, channel reciprocity manifests itself by the similarity of the overall shapes between the pair. By plotting smoothed points, conformal information about the overall shape is extracted despite the local variations. In our algorithm, we adopt Locally Weighted Scatter Plot (Lowess) smoothing \cite{Clev:1979}, a curve fitting method that calculates the smoothed value by applying locally weighted regression over a span. Fig.~\ref{fig:smoothed} depicts two PSD curves obtained by two communicating wireless nodes and their corresponding curves after applying Lowess smoothing with a span of 0.4. From Fig.~\ref{fig:smoothed}, we can see that the Lowess curves coincide with each other almost exactly and the overall shapes are preserved, even though the original ones differ from each other in most of the locations. 
\subsection{Curve Encoding}\label{sec:curve_encoding}
By using curve smoothing, we obtain two highly similar curves. 
To solve the encoding problem, let us first briefly consider several alternative methods: 1)~encode in accordance with an approximation function that describes the curve; 2)~encode in accordance with the statistical properties of the curve; 3)~encode by describing the shape of the response. We adopt the third one for the following reason. As mentioned in Section~\ref{sec:curve_smoothing}, channel reciprocity is readily seen by the similarity of the overall shapes between curves. Hence, encoding by describing the shape should preserve most of the information shared by the two ends. By way of contrast, extracting secrets from the statistical properties definitely suffers from losing much of the mutual information. And the approximation function does not tolerate even small deviations, but measurement error and interference make such deviations quite common.
\begin{figure}[t]
  			\centering 
  				\includegraphics[width=0.8\columnwidth]{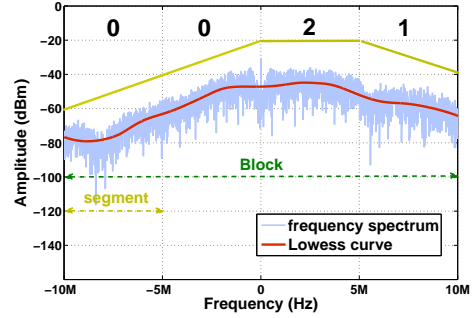}
  			\caption{An example of curve encoding.} 
  			\label{fig:coding}
  		\end{figure} 		
Fig.~\ref{fig:coding} gives an example of curve coding. The curve obtained in a certain band is treated as a block, which can be divided into varying number of segments of equal length, and then the segments are mapped to one of three curve patterns which are of the same length, as shown in Fig.~\ref{fig:coding}. These three patterns are indexed as 0, 1, and 2. The three ``predetermined'' patterns describe the ascending, descending and steady trend of the curves respectively. By ``predetermined'', we mean that the indices and the shapes of the patterns are well known to all wireless nodes. The gradient of the ascending and descending lines, however, is decided by each node according to the maximum and minimum values of the smoothed curve, and the length of the segment. We have designed that pattern generation thus to tolerate measurement errors and different device settings. For example, two communicating nodes may wish to use different tx/rx gains that would amplify the signals differently. Since each pattern is related to the locally received signals, it describes the shape correctly without the need to negotiate with the other node. We set the gradient of the ascending pattern to be relative to $\frac{max-min}{\text{\# of samples in each segment}}$, and likewise for the descending pattern is relative to $-\frac{max-min}{\text{\# of samples in each segment}}$. The segment is then mapped to the most similar of the three patterns by measuring the discrete Fr\'echet distance \cite{Wien94computingdiscrete} $\delta_{dF}$ between the segment and the patterns, which measures the similarity of two polygonal curves while taking the location and ordering of the points along the curves into consideration. The smaller the distance, the more is the similarity the two curves share. 
The complete algorithm is presented in Algorithm~\ref{alg:coding} and Algorithm~\ref{alg:pattern}.

  \section{Experimental Validation}\label{validation}
In this section we study four important metrics to measure the performance of Puzzle. 
\vspace*{-4mm}
\begin{itemize}
	\item  \textbf{Entropy}:
	 Entropy measures the unpredictability of a random variable X. It is defined as
	\[H(X)\ =\ -\sum_{i=1}^{n}p(x_{i})\log_{2}p(x_{i})\]
	where ${x_{1},\cdots,x_{n}}$ are possible values of $X$. 
	
	\item \textbf{Bit Mismatch Rate}:
	 Bit mismatch rate is defined as the ratio of the number of bits between Alice and Bob that do not match and the number of bits extracted from the shape of the spectrum.
	
	\item \textbf{Correlation}:
	Correlation $\rho_{x,y}$ is defined as
		$$\rho_{x,y} = \frac{\sum\limits_{i=1}^n (x_{i}-\bar{x})(y_{i}-\bar{y})}{\sqrt{\sum\limits_{i=1}^n  (x_{i}-\bar{x})^{2}\sum\limits_{i=1}^n (y_{i}-\bar{y})^{2}}}$$ We use correlation to measure the dependence of codes generated by Puzzle relative to different distance between Bob and Eve.
	
	\item \textbf{Leakage}:
	 Letting $p_{mis}$ be the mismatch rate between Alice and Eve, we define the leakage between them as
		\[ 
		leakage =
			\begin{cases}
				1-\frac{p_{mis}}{0.5} & \text{if\ $p_{mis}\ <\ 0.5$}\\
				0 &  \text{otherwise}
			\end{cases}
		\]
\end{itemize}

 

\subsection{Environment and System}
\label{sub:environment}

The measurement environment is a lab where there are 6 cubicles. Data were collected during daytime (from 7:00 am to 6:00 pm). 
Human activities introduced a certain level of interference in the channel, but generally speaking, 
the environment is quite stable. 
We conducted the experiment in such a stable environment because we wanted to see clearly the performance comparisons without risking mismatches caused by the changes of the channel itself. In theory, further implementation in mobile environment would give both higher mismatch rate and higher secret bit extraction rate.

The communication system consists of three software-defined transceivers. Each of their RF chains contains an XCVR2450 (RF front end), a NI-5781 (data converter module) and an NI PXIe-7965R (a Xilinx Virtex-5 FPGA). Two of the three transceivers transmit at 2.45 GHz with 20MHz bandwidth. We call these two transceivers Alice and Bob. The third transceiver, Eve, overhears the communication. During reception, each transceiver records the I and Q samples at a sampling rate of 100 MHz and down converts to the baseband. The received samples are then sent to the NI PXIe-8133, an RTOS-based controller, through two direct-memory-access (DMA) channels, which have a data streaming rate that is as high as 800 MB/s. Except for the experiment done in Section~\ref{exp:CSI}, all the results of Puzzle are obtained based on the PSD of 10240 received samples with QPSK modulation.  

\subsection{Performance Evaluation}
\subsubsection{Entropy and mismatch rate}\label{exp:CSI}
We first compare Puzzle with the frequency domain secret key generation method with 2-bit quantization \cite{freq}, which in the rest of this paper we refer to as the CSI-2bit. We choose CSI-2bit as the basis for bit mismatch rate and entropy comparison because, to the best of our knowledge, it achieves the highest bit generation rate along with a low mismatch rate. Coarse-grained method like RSS-based ones achieve only 1$\sim$3 bits per packet. We conducted an experiment where packets were transmitted over coherence time using OFDM in a 20MHz band, with each OFDM symbol consisting of of 72 subcarriers. A channel frequency response is extracted from each OFDM subcarrier. The same channel frequency response was used in both Puzzle (to construct curves) and CSI-2bit (to quantize the response). By dividing the curve composed of the 72 channel frequency responses into a certain number of segments of even length for Puzzle, and by selecting a certain number of frequency responses evenly from all the 72 subcarriers for CSI-2bit, we extracted the respective secrets from each packet for the two methods, thus obtaining secrets of different lengths. No device calibration is done as it is orthogonal to the secrecy metrics we wish to compare. 

\begin{figure}[!htbp]
\centering
	
	\begin{minipage}[t]{\columnwidth}
	\begin{subfigure}[t]{.49\columnwidth}
	\centering
	\includegraphics[width = \columnwidth]{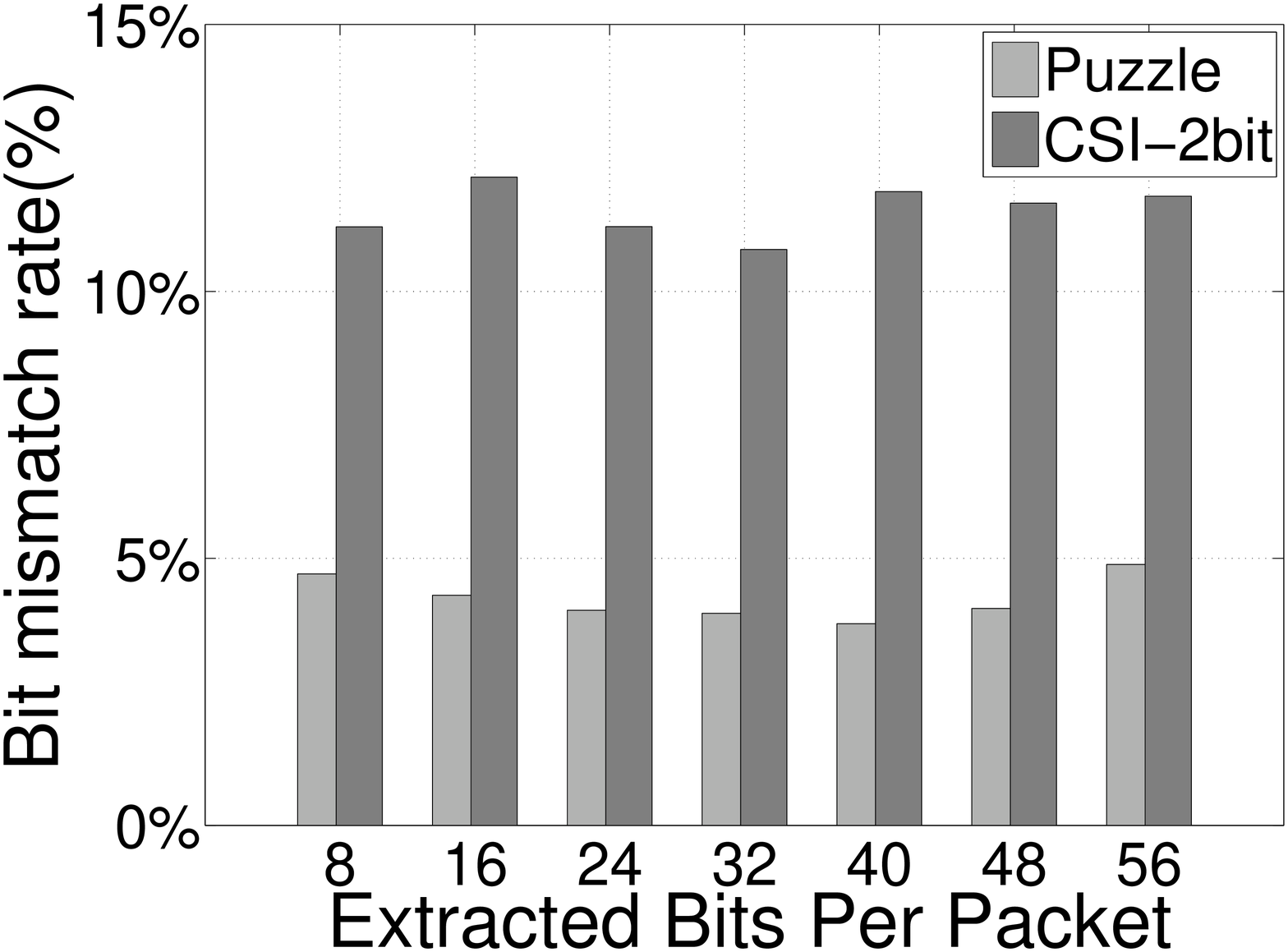}
	\caption{ Mismatch rate of different bit generation rates.} 
	\label{fig:MisMatch}
	\end{subfigure}
	\begin{subfigure}[t]{.49\columnwidth}
			\centering
			\includegraphics[width = \columnwidth]{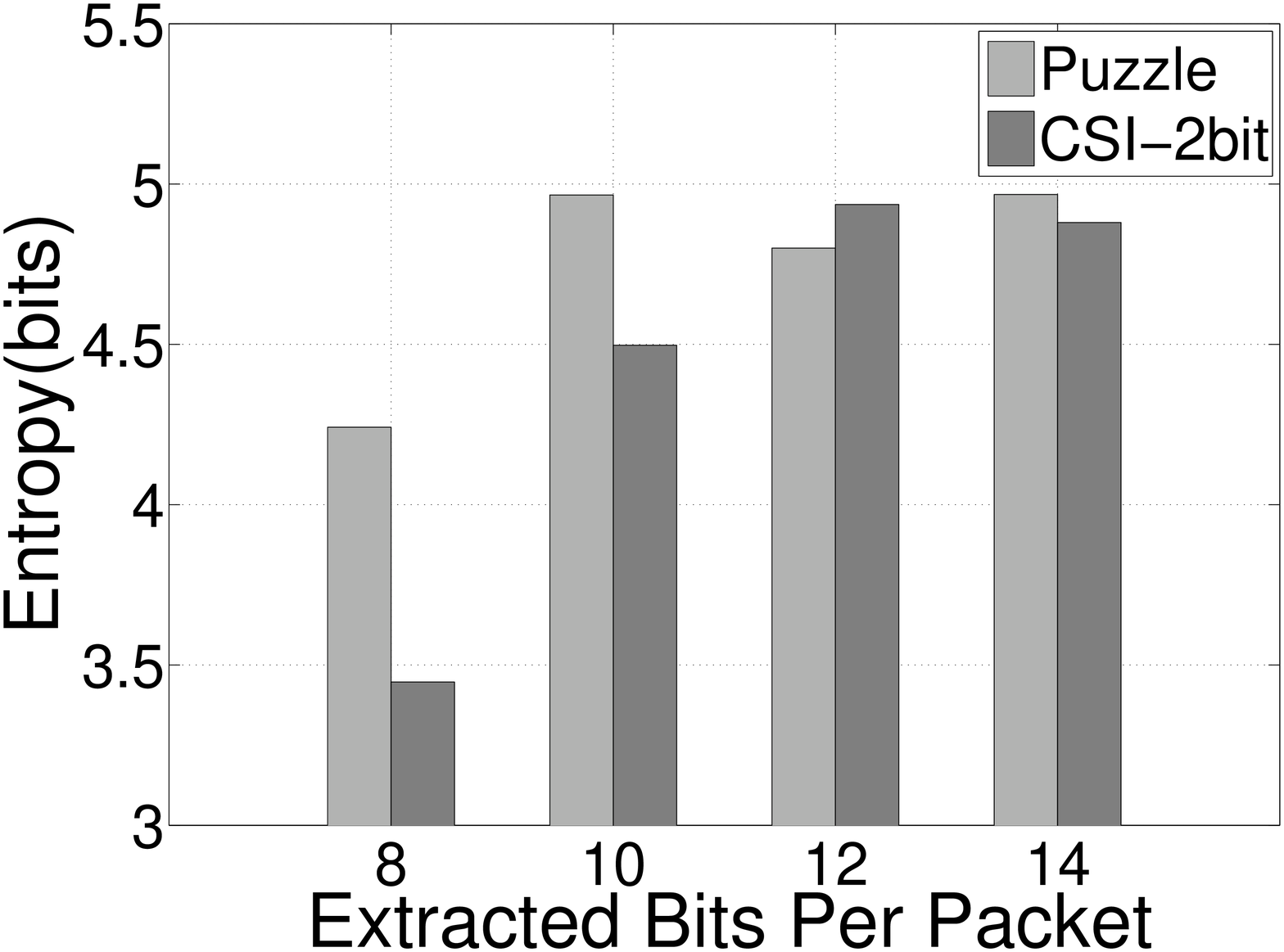}
			\caption{ 
			Puzzle produces a comparable amount of entropy as CSI-2bit.}
			\label{fig:entropy}
		\end{subfigure}
		\vspace{0\baselineskip}
	\caption{Bit Mismatch Rate and Entropy}
	\end{minipage}
	
\end{figure}

Fig.~\ref{fig:MisMatch} shows that Puzzle outperforms CSI-2bit in bit mismatch rate for bit generate rates from 8bit/pkt to 56bit/pkt. On average, Puzzle has a 63\% lower bit mismatch rate than CSI-2bit. It is worth noting CSI-2bit has an option of online device calibration but that procedure requires the two communicating nodes collect CSI over hundreds of coherence intervals, therefore it has high overhead and is not practical for fast secret sharing.

From Fig.~\ref{fig:entropy}, we first see that Puzzle produces a comparable amount of entropy as CSI-2bit does. It implies that shape-based method does not harm the entropy compared to quantization-based method. Another thing we notice is that, the entropy of the generated bits does not increase linearly with the number of bits used to encode them. This is caused by the fact that neighboring subcarriers are correlated. Therefore, the claim made by Liu et al \cite{freq} that, CSI-2bit can generate 60-bit secret per packet by using 30 subcarriers in a 20MHz band, is not accurate. Fig.~\ref{fig:entropy} shows that for a 14-bit code generated by Puzzle or CSI-2bit, the real secret contained in it, is not longer than 5-bit. And the entropy is saturating as the bit generation rate increases.

\subsubsection{Correlation of codes relative to distance}
\begin{figure}[!t]
\centering
\begin{minipage}[t]{\columnwidth}
	\begin{subfigure}[t]{.49\columnwidth}
	\centering
	\includegraphics[width = .95\columnwidth]{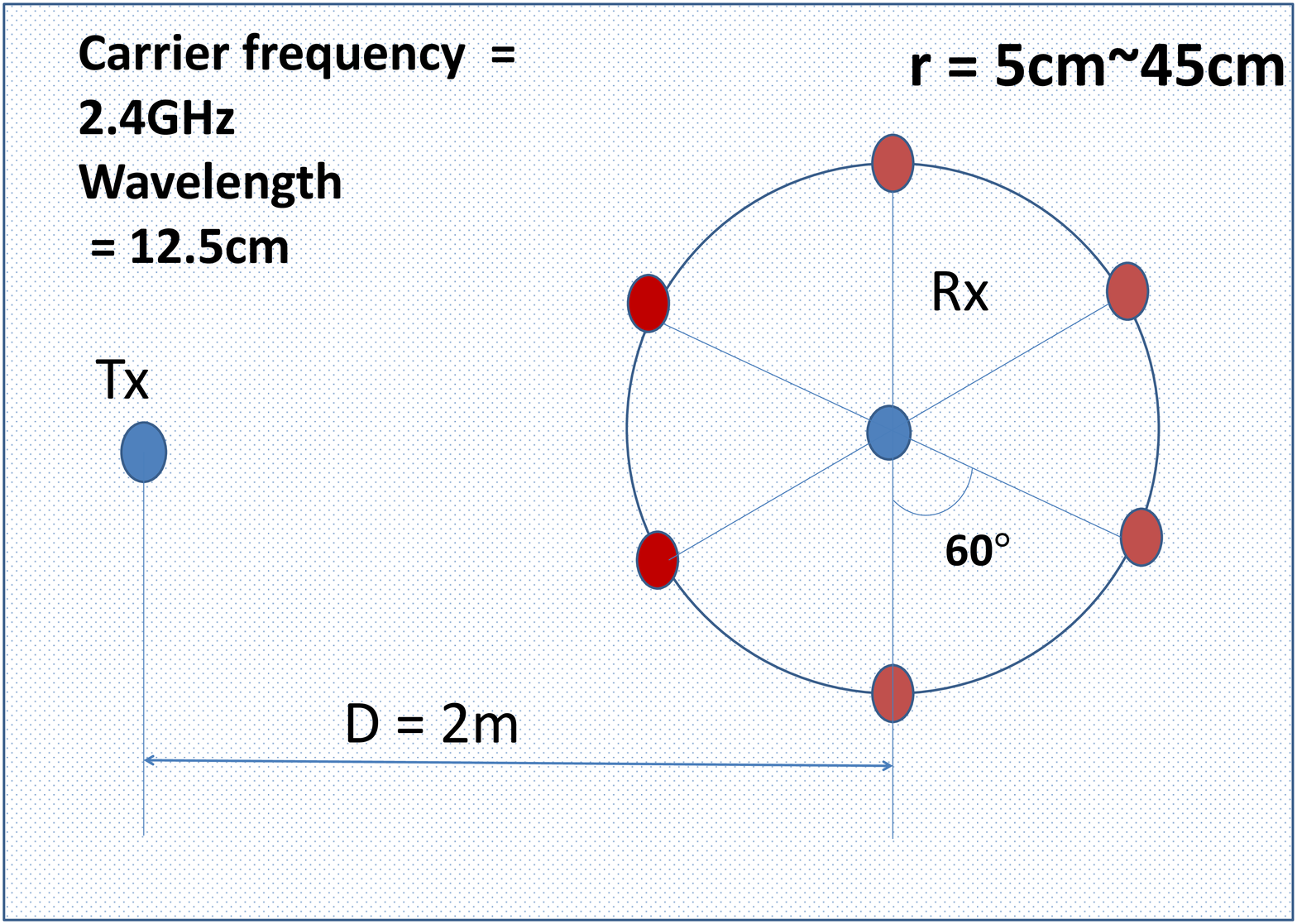}
	\caption{ Deployment of correlation experiment.}
	\label{fig:deployment}
\end{subfigure}
\begin{subfigure}[t]{.49\columnwidth}
	\centering
	\includegraphics[trim = 0mm 0mm 0mm 0mm, clip, width = \columnwidth]{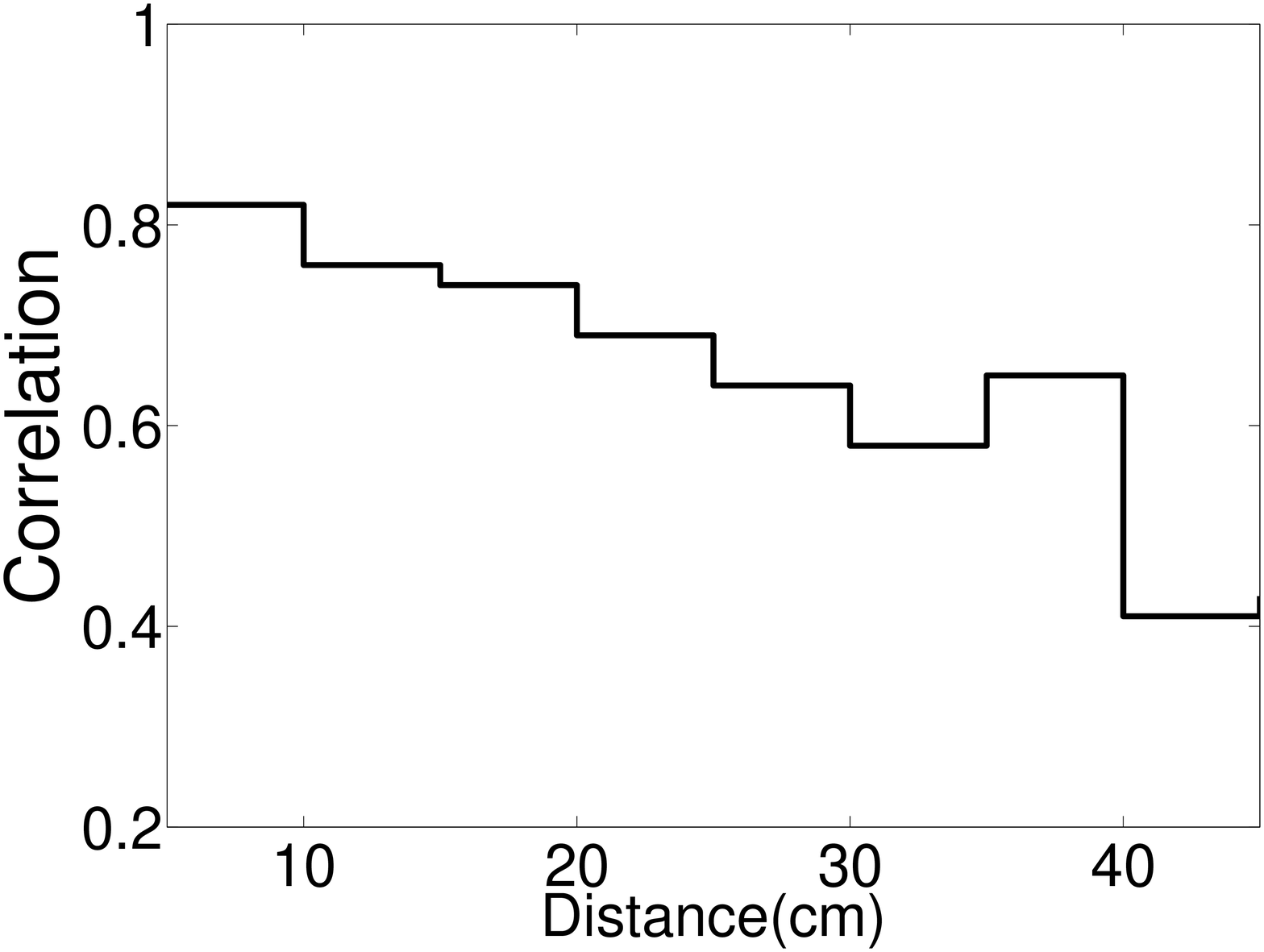}
	\caption{ The correlation of two codes generated by Puzzle relative to distance.}
	\label{fig:correlation}
\end{subfigure} 
\caption{Deployment and result of correlation experiment}
\end{minipage}
\end{figure}

To evaluate the resistance to an eavesdropping attacker, we establish the correlation of bits generated by two receivers at different distances. We performed an experiment where we fixed the distance between one transmitter and one receiver, and then placed another receiver at a certain distance away from the first receiver along 6 orientations as shown in Fig.~\ref{fig:deployment}. Each frequency response curve is segmented into 4 pieces. We measured the correlation between the codes produced by the two receivers at distances ranging from 5cm away to 45cm away. To be more specific, we measured the correlation of 6 pairs of locations by fixing the first receiver and moving the second one $60^{\circ}$ apart at each distance. Figure~\ref{fig:correlation} shows the result. We see that the correlation decreases rapidly as the distance between two receivers increases. In practice, it is reasonable to assume that eavesdroppers are beyond one meter away, otherwise they suffer from high risk of exposures. Therefore Puzzle is robust against eavesdropping.

\label{fig:correlationexp}

\begin{figure}[h]
\centering 
 \begin{subfigure}[t]{\columnwidth}
 \centering
  \includegraphics[trim=0cm 1.5cm 0cm 0cm, clip=true, width=.85\columnwidth]{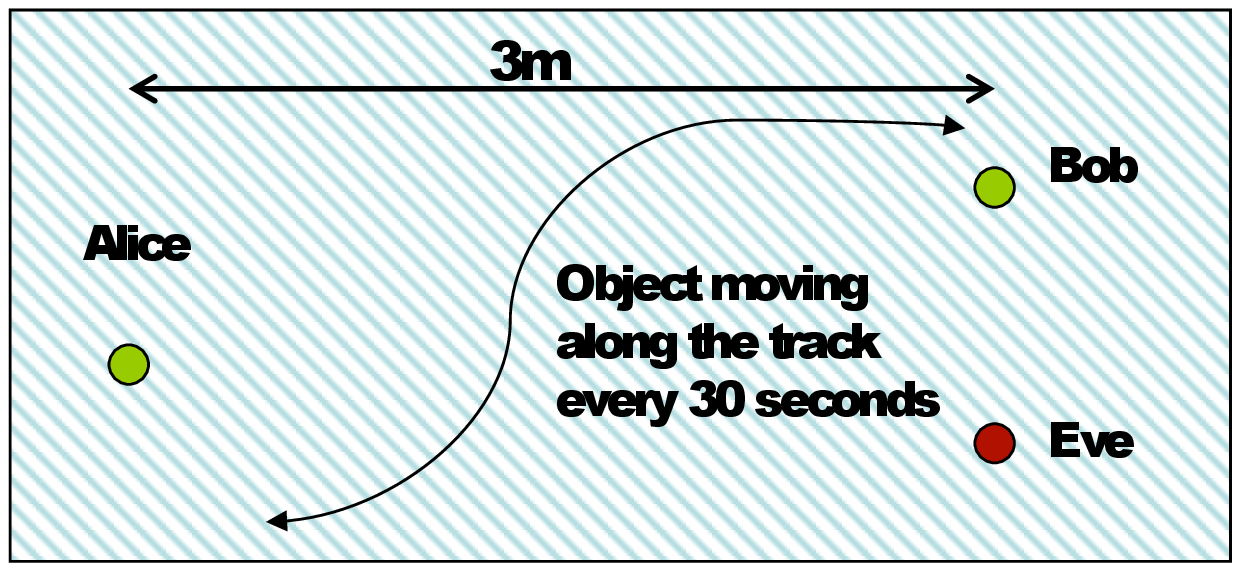}
  \caption{An object moves between Alice and Bob with a certain temporal pattern and Eve 
  overhears the transmission from Alice to Bob.}
    			\label{fig:MovingObj}
\end{subfigure}	

\begin{minipage}{\columnwidth}
\begin{subfigure}[t]{.49\columnwidth}
	\centering
	\includegraphics [trim = 0.5mm 1mm 0.5mm 0.5mm, clip, width=\columnwidth]{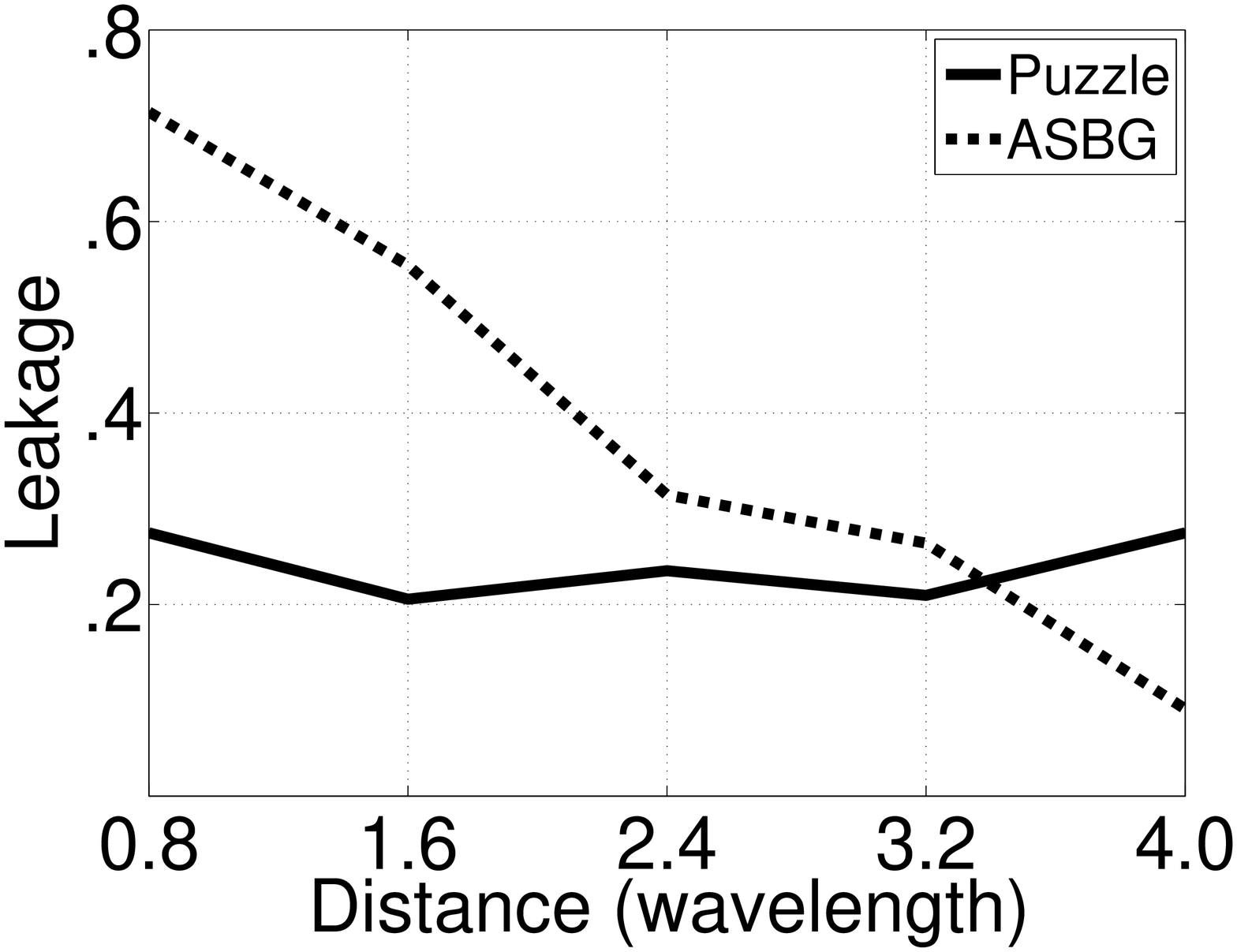}
	\caption{Leakage relative to distance between Bob and Eve. Puzzle has a stable low leakage rate irrespective of the distance.}
	\label{fig:subfigLeakage}
\end{subfigure}
\begin{subfigure}[t]{.49\columnwidth}
	\centering
	\includegraphics [ trim = 0.5mm 1mm 0.5mm 0.5mm, clip, width=\columnwidth]{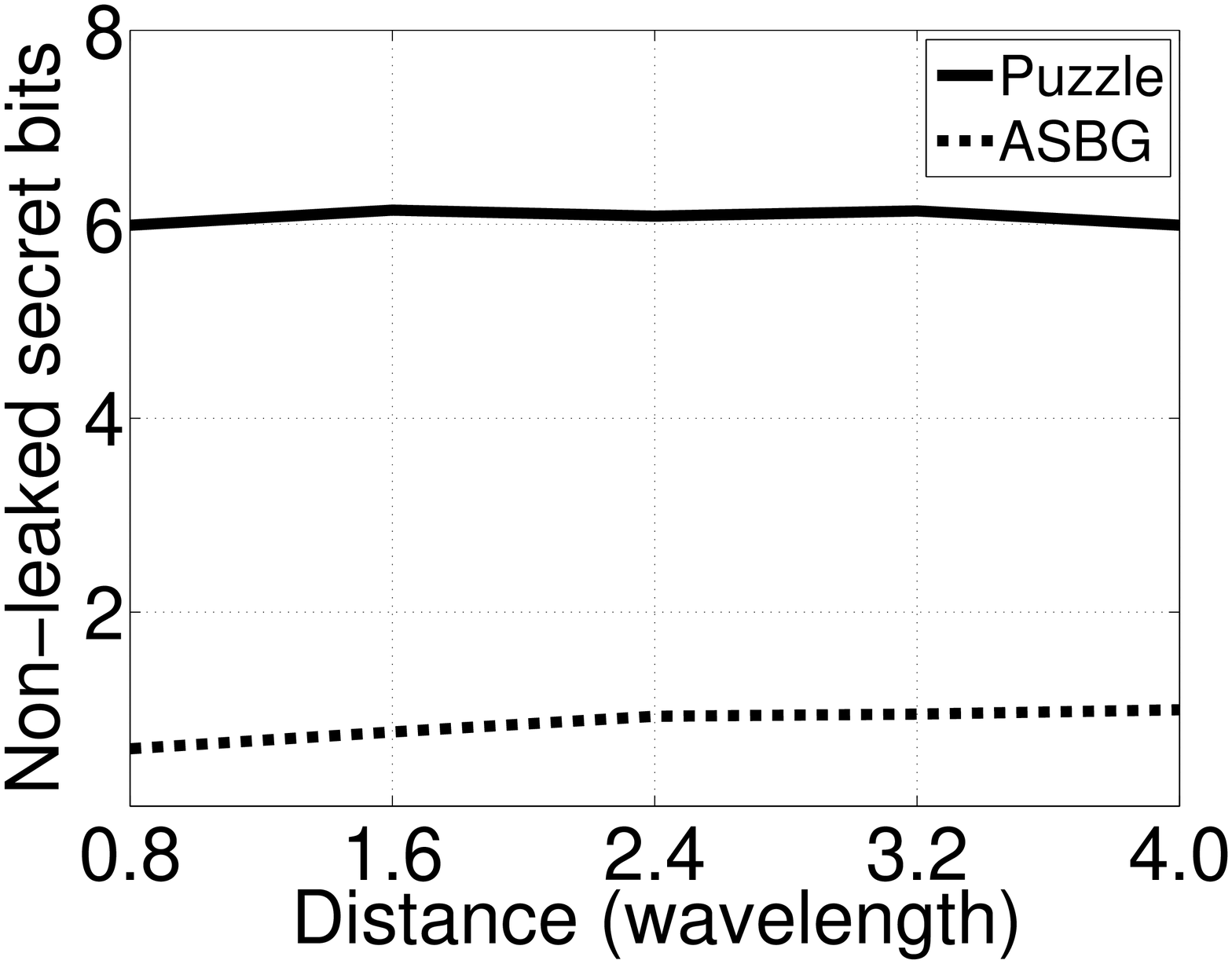}
	\caption{Non-leaked secrete bits produced by Puzzle and ASBG per packet, relative to the distance between Bob and Eve.}
	\label{fig:subfigLeakEntropy}
\end{subfigure}
\caption{Performance: Leakage}\label{fig:leakperformance}
\end{minipage}
\end{figure}
\subsubsection{Leakage}\label{sub:comparison}
Towards validating the resistance to the planned movement attacker (cf.~Section~\ref{model}), we 
compared the leakage performance of a the state-of-the-art RSS-based method ABSG and Puzzle by moving an object across the transmission path between Alice and Bob, while placing an eavesdropper near Bob, as shown in Figure~\ref{fig:MovingObj}. 
Since ABSG like many other RSS-based methods asks the two communicating ends to drop some RSS values based on certain thresholds and to exchange the indices of those values, Eve knows exactly which RSS probe is used by Bob but dropped by herself. In this case, we assume that Eve makes a random guess as to the quantization result with a success rate of 50\%. We calculate the mismatch rate of Eve's and Bob's bits to be the combination of the actual mismatch rate between them and the failure rate of the random guess. And again, we segment the frequency response curves into four pieces.

Fig~\ref{fig:subfigLeakage} shows the leakage of our algorithm against that of ABSG over a distance from $10$ cm to $50$ cm. It is clear that Puzzle is much more insensitive to the threat of planned movement. Furthermore, due to the fact that Puzzle has a much higher secret generation rate ($4*log(3) \approx 6.3$ bits/pkt) than ASBG (1 bits/pkt), the non-leaked secret produced by Puzzle is much larger. Fig~\ref{fig:subfigLeakEntropy} shows the result. It is worth noting that although 4 wavelength might not sound like a large distance in practice, our blocking objects are not large either. The variations induced by larger obstacles, like a train passing by or the example mentioned above, might impact a much larger distance in practice.

  

\setlength{\bibsep}{0pt plus 0.2ex}
{\small \bibliographystyle{acm}
  \bibliography{citations}}
\end{document}